\documentstyle[12pt,aasms4]{article}
\slugcomment{Submitted to ApJ}
\lefthead{Perea et al.}
\righthead{Non-linear Dependence of $L_B$ on $L_{FIR}$ and $M_{H_2}$
among Spiral Galaxies and Effects of Tidal Interaction}

\newcommand{\mdos}{$\rm M_{H_2}$}
\newcommand{\fir}{$\rm L_{FIR}$}
\newcommand{\lum}{$\rm L_B$}

\def\kms  {\ifmmode{{\rm ~km~s}^{-1}}\else{~km~s$^{-1}$}\fi}

\newcommand{\pepe}{\multicolumn{1}{c}{---}}
\newbox\grsign \setbox\grsign=\hbox{$>$} \newdimen\grdimen 
\grdimen=\ht\grsign
\newbox\simlessbox \newbox\simgreatbox
\setbox\simgreatbox=\hbox{\raise.5ex\hbox{$>$}\llap
    {\lower.5ex\hbox{$\sim$}}}\ht1=\grdimen\dp1=0pt
\setbox\simlessbox=\hbox{\raise.5ex\hbox{$<$}\llap
     {\lower.5ex\hbox{$\sim$}}}\ht2=\grdimen\dp2=0pt

\begin{document}
\title{Non-linear Dependence of $L_B$ on $L_{FIR}$ and $M_{H_2}$ 
among Spiral\\ Galaxies and Effects of Tidal Interaction}
\author{J. Perea, A. del Olmo, L. Verdes-Montenegro}
\affil{Instituto de Astrof\'\i sica de Andaluc\'\i a, CSIC, Apdo.~3004,
18080 Granada, Spain.\\ ({\it jaime@iaa.es, chony@iaa.es, lourdes@iaa.es})}
\and 

\author{M.~S.~Yun}
\affil{National Radio Astronomy Observatory\footnote{The National Radio
Astronomy Observatory is a facility of the National Science Foundation
operated under cooperative agreement by Associated Universities, Inc.},
P.O. Box 0, Soccoro, NM~~87111 ({\it myun@nrao.edu})}

\received{22 Apr 1997}
\accepted{25 Jun 1997}
\begin{abstract}

Through the study of a carefully selected sample of isolated spiral
galaxies, we have established that two important global physical
quantities for tracing star forming activities, $L_{FIR}$ and
$M_{H_2}$, have non-linear dependence on another commonly cited global
quantity $L_B$. Furthermore we show that simple power law relations
can effectively describe these non-linear relations for spiral
galaxies spanning four orders of magnitude in FIR and \mdos\ and
nearly three orders of magnitude in $L_B$.  While the existence of
non-linear dependence of $M_{H_2}$ (assuming a constant CO-to-H$_2$
conversion) and $L_{FIR}$ on optical luminosity ($L_B$) has been
previously noted in the literature, an improper normalization of
simple scaling by $L_B$ has been commonly used in many previous
studies to claim enhanced molecular gas content and induced activities
among tidally interacting and other types of galaxies.  We remove
these non-linear effects using the template relations derived from the
isolated galaxy sample and conclude that {\it strongly interacting
galaxies do not have enhanced molecular gas content}, contrary to
previous claims.  With these non-linear relations among $L_B$,
$L_{FIR}$ and $M_{H_2}$ properly taken into account, we confirm again
that the FIR emission and the star formation efficiency
($L_{FIR}/M_{H_2}$) are indeed enhanced by tidal interactions.  Virgo
galaxies show the same level of \mdos\ and \fir\ as isolated
galaxies. We do not find any evidence for enhanced star forming
activity among barred galaxies.

\end{abstract}


\section{Introduction}

The molecular gas (H$_2$) and far-infrared (FIR) contents are
important quantities in analyzing the star formation activity in
galaxies.  They have been analyzed in the literature for different
galaxy samples (\cite{ys91} and references therein,
\cite{sof93,bc93,y96,sol97}) although largely among the IR luminous
galaxies. These two quantities correlate linearly assuming a constant
CO-to-H$_2$ conversion factor, and massive star formation within
molecular clouds is sufficient to explain the observed trend in most
cases. On the other hand, several authors have noted that these
quantities do {\it not} correlate linearly with the total blue
luminosity (L$_B$) in the sense that $M_{H_2}$/L$_B$ and
L$_{FIR}$/L$_B$ increase with increasing L$_B$ (\cite{y87,y89,ss89}).
Nevertheless, a simple normalization using L$_B$ has been commonly
used to claim enhanced activities among different groups of galaxies
(e.g., \cite{bc93,co94,dh97}) while what they have actually shown is
this residual dependence on the luminosity.  In this paper, we first
examine the nature of non-linearity involving L$_B$, and then
re-examine the previous claims of enhanced molecular gas content and
star forming activity based on the normalization by L$_B$. We show
here that the enhanced molecular gas content among bright interacting
galaxies reported in the literature is entirely due to the
non-linearity effect, but the excess in the FIR emission is real and
is largely due to real enhancement induced by environmental effects.

We have compiled from the literature an extensive comparison sample of
207 galaxies in varying interaction classes and environment (isolated,
weakly perturbed, strongly perturbed, Virgo cluster), representing a
wide range of optical luminosity (10$^{8.6}~L_{\odot}$ $<$ L$_B$ $<$
10$^{11.4}~L_{\odot}$) and for which FIR, CO, and optical luminosity
(as measured by the blue luminosities \lum ) are available.  The FIR
luminosity is obtained as $log(L_{FIR}/L_{\odot})=log(FIR) + 2 log(D)
+ 19.495$ where D is distance in Mpc and $FIR=1.26 \times 10^{-14}
\times (2.58 I_{60} + I_{100})$ W~m$^{-2}$ ({\cite{h88}). The 60
$\mu$m and 100 $\mu$m fluxes are obtained from the NED\footnote{The
NASA/IPAC Extragalactic Database (NED) is operated by the Jet
Propulsion Laboratory, California Institute of Technology, under
contract with the National Aeronautics and Space Administration.}
database. The molecular hydrogen mass is derived as
$M_{H_2}=4.82~\alpha I_{CO} d^2_B~M_{\odot}$, assuming a constant
CO-to-H$_2$ conversion factor of $\alpha$ = $N_{H_2}/I_{CO} = 3\times
10^{20}$ cm$^{-2}$ (K \kms)$^{-1}$, where $I_{CO}$ is the velocity
integrated CO intensity in K \kms\ and $d_B$ is the half-power beam
diameter in pc at the source distance (\cite{s91}). The use of a
constant conversion factor is discussed further in $\S$ 2.  Because
many nearby galaxies typically subtend several arcmin in size and are
much larger than the beam of the telescopes used, the comparison data
consist mostly of the CO surveys using at least partial mapping. The
optical luminosity is derived from the $\rm B^0_T$ magnitude from the
RC3 catalogue, corrected for galactic absorption (using the extinction
value given by \cite{bh}, with the reddening law from \cite{sm}) and
internal extinction (\cite{dv} RC3) using the redshift given in
NED. We adopted a value for the Hubble constant of $\rm H_0\, = 75~km
\, s^{-1} \, Mpc^{-1}$.

\section{Relationship among $L_B$, $L_{FIR}$ and $M_{H_2}$ for
Isolated Galaxies}

In order to first understand the intrinsic non-linear nature of $L_B$
with respect to FIR luminosity and molecular gas content, we have
assembled from the literature a reference sample of isolated galaxies
representing a wide range of luminosity that extends from L$_B$ =
$10^{8.6}$ to $10^{10.9}~L_{\odot}$.  Our isolated galaxy sample
consists of 68 objects from the distance limited survey of the Nearby
Galaxies Catalog (\cite{t88}) by \cite{s93} and the class 0 objects of
the IRAS selected sample by \cite{ss88}. For six galaxies only upper
limits are available on the CO luminosity. Morphological types range
from Sa to Sd galaxies, each distributed along the entire luminosity
range, and the types around Sc are the most abundant. This sample
lacks completeness because the galaxies studied in the literature are
frequently biased towards infrared luminous galaxies. On the other
hand, our isolated galaxy sample is sufficient for this comparison
study in the sense that the range of optical luminosity present
in other comparison samples are basically included.

In Fig. 1a, we plot \lum\ against \mdos\ for this sample. The data can
be well described by a simple power law relation
\begin{eqnarray}
{log L_B} & = & (0.57\pm 0.03)~log M_{H_2} + (4.9 \pm 0.6) 
\end{eqnarray}
\noindent The coefficients are determined by a linear regression
taking into account the upper limits using the Astronomy Survival
Analysis package (ASURV\footnote{Astronomy Survival Analysis (ASURV)
Rev 1.2 package is a generalized statistical analysis package which
implements the methods presented by Feigelson \& Nelson (1985) and
Isobe et al. (1986) and are described in detail in Isobe \& Feigelson
(1990) and La Valley et al. (1992)}). We do not find any dependence on
the morphological types. Sage \& Solomon (1989) report a slope of 0.53
$\pm$ 0.07 for a sample of isolated and weakly perturbed galaxies, and
Young et al. (1989) find a slope of 0.72 $\pm$ 0.03 for a sample of
124 galaxies with CO data.  Removing the 5 nearby dwarf galaxies with
lowest luminosities (L$_B$ $<$ 7 $\times$ 10$^{7}$ L$_{\odot}$) that
clearly separate from the overall tendency, we find a slope of 0.54
for the Young et al. sample -- which is consistent with our result.

In Fig. 1b we plot \fir\ against \mdos, whose ratio is a measure of
the star formation efficiency (SFE) (see \cite{ys91}). This
correlation has been widely considered in the literature and found to
be nearly linear when a large range of luminosities are considered,
\begin{eqnarray}
{log L_{FIR}} & = & (0.90 \pm 0.05)~log M_{H_2} + (1.5 \pm 0.4) 
\end{eqnarray}
\noindent with a similar dispersion as that of L$_B$-M$_{H_2}$ relation.

These plots demonstrate that L$_B$ increases more slowly compared with
the molecular gas content (for a constant CO-to-H$_2$ conversion
factor) and far-infrared luminosity, and more importantly, these
non-linear dependencies are present even among these carefully
selected isolated spirals.  While the physical cause for the observed
non-linear relations is uncertain, it is suggestive that the physical
mechanism responsible for the FIR and CO ($M_{H_2}$) emission may be
closely related, but they are distinct in nature from optical
emission.  For the ease of the subsequent discussions, we have
parameterized the above empirical non-linear relations in terms of
M$_{H_2}$, but we do not claim that M$_{H_2}$ (or any of the three
physical quantities compared here) is the fundamental driver.

If one postulates that the root of the non-linear relations shown in
Eqs. 1 \& 2 lies in the FIR and CO emission, then an obvious possible
cause is the metallicity dependence on the luminosity.  In fact, a
positive correlation between metallicity and absolute magnitude has
been obtained by Roberts \& Haynes (1994) for a large sample of spiral
galaxies, and Arimoto et al. (1996) have proposed an empirical
dependence between CO--to--H$_2$ conversion and absolute magnitude
based on their observational data and the CO emission model of
Sakamoto (1996).  Indeed there are clear indications that the CO
intensity may not be a good tracer of the molecular content among
metal--poor galaxies (\cite{coh88,rub91}).  Also, the non-linear
dependence in Eq. 1 is nearly reproduced by the metallicity dependence
derived by Roberts \& Haynes if the CO emission is assumed to depend
linearly on the metallicity.  However, there are observational and
theoretical reasons to believe that such a metallicity
dependence cannot work.  For example, the observational evidence for
metallicity dependence offered by Arimoto et al. becomes marginal if
the low spatial resolution data on M51 and LMC are removed.  Also,
full chemical and radiative transfer modeling of CO emission by
Wolfire, Hollenbach, \& Tielens (1993) and Maloney \& Wolfire (1997)
suggests that CO emission is a robust tracer of $M_{H_2}$ against
metallicity variation -- no metallicity dependence is expected until
the metal abundance becomes low enough to make the CO transition
optically thin.

Alternatively if the non-linear relations derived above are rooted in
L$_B$, then optical extinction and scattering model may offer an
explanation.  A naive expectation is that larger extinction in more
massive galaxies with larger dust column density would result in a
slower increase in \lum\ with size (e.g., Young et al. 1989).  Buat \&
Xu (1996) also find a good correlation between the optical depth in B
and the ratio of FIR to blue luminosity. Therefore, it is natural to
suspect that the root of the observed non-linearity may lie with $L_B$
rather than with $L_{FIR}$ or $M_{H_2}$.

In a simple geometrical extinction model in which optically thick
tracers such as blue light represent surface area and optically thin
tracers such as FIR and integrated CO intensity represent volume
elements, a non-linear dependence is naturally expected between these
two classes of tracers -- i.e., $L_B \propto [M_{H_2}]^{2/3}$ for a
spherical uniform cloud.  In reality, the structure of the ISM is more
complex and clumpy or filamentary, and the effective optical depth is
significantly reduced compared with a uniform screen model by
extinction geometry and efficient scattering (Witt et al. 1992,
1996). Nevertheless, it is remarkable that the non-linear relation
between $L_B$ and $M_{H_2}$ is observed with nearly the same power law
dependence as predicted by this simple model, and one may infer that
the basic assumptions of the model stated above must hold at the
clump-interclump size scales. We note that the predicted exponent of
2/3 is independent of the cloud shape as long as the cloud structure
remains self-similar at different size scales.

The extinction model also correctly predicts other optical depth
dependent trends. For example, the exponent of the correlation between
$L_B$ and $L_{FIR}$ is expected to be {\it steeper} than that of
$L_B$--$M_{H_2}$ because additional surface element provided by HI
among the low luminosity galaxies is partly compensated in FIR by the
dust emission associated with the HI gas.  A formal fit for the
$L_B$--$L_{FIR}$ relation in our isolated galaxy sample gives
\begin{eqnarray}
{log L_{B}} & = & (0.65 \pm 0.09)~log L_{FIR} + (3.9 \pm 0.9) 
\end{eqnarray}
\noindent and the slope of the correlation is indeed larger than that
of $L_B$--$M_{H_2}$ ($0.57\pm0.03$).  We predict that the optical
luminosity at longer wavelengths ($L_H$ or $L_K$) would produce a more
linear relationship with $M_{H_2}$ than $L_B$.

\section{Re-evaluation of the Tidal and Environmental Effects}

Regardless of their physical causes, two global physical quantities
$L_{FIR}$ and $M_{H_2}$ show clearly non-linear dependence with
respect to $L_B$ among a large sample of isolated galaxies. The
relations are well described by a simple power law spanning four
orders of magnitude in $L_{FIR}$ and \mdos\ and nearly three orders of
magnitude in $L_B$ (see $\S2$).  Therefore, any comparison of
$L_{FIR}$ and $M_{H_2}$ normalized simply by $L_B$ produce a
misleading result.  This is particularly true for the studies of
bright interacting galaxies found in the literature where the samples
are often selected based on their high luminosity.  Here we
re-evaluate the impact of tidal and environmental effects on
individual galaxies by examining their global properties using the
empirical relations derived from our isolated galaxy sample as
templates in order to discriminate the size--dependent effects from
the environmentally induced effects.  Deviations from the template
power laws (Eqs. 1 \& 2) are evaluated below for galaxies suffering
different degrees of interactions. We also examine whether the
presence of a bar correlates with any enhancement in the star
formation efficiency.  While each subsamples analyzed here represent
different range of luminosity with different mean luminosity, almost
the entire range of optical luminosity considered are represented in
our isolated galaxy sample, and the use of these template relations
are justified.

Both the FIR and CO data available in the literature are severely
sensitivity limited, and a special care has to be made to incorporate
the upper limits properly in the analysis.  Standard statistical tests
such as Kolmogorov--Smirnov (KS) or Mann--Whitney cannot be rigorously
applied to these heavily censored data, and we instead use equivalent
tests provided by ASURV. In particular to quantify the statistical
probability of the compared samples representing the same distribution
function, we use the logrank and Peto--Prentice generalized Wilcoxon
test which are known to be the most robust against the censoring
pattern.

\subsection{Interacting and Cluster Galaxies}

The entire sample studied here consists of 139 galaxies, divided into
three subclasses: weakly perturbed (WP), strongly perturbed (SP), and
Virgo Cluster (VC) galaxies. The WP sample has 43 galaxies including
interaction class 1, 2 \& 3 of \cite{ss88} and class 2 objects from
the luminous IRAS sample by Sanders et al. (1991). The SP sample has
38 galaxies including interaction class 4 of \cite{ss88}, interaction
class 3 \& 4 of Sanders et al. (1991), and closely interacting pairs
from Combes et al. (1994). The definitions of ``weakly perturbed'' and
``strongly perturbed'' are given in the references listed above, SP
galaxies being those identified as the final stages of mergers. In
addition, we have constructed a sample made of 58 member galaxies in
Virgo cluster, including both bright (Kenney \& Young 1988ab) and
faint spirals (Boselli et al. 1995). In the SP and VC sample, 3 and 18
galaxies have only upper limits in CO, respectively.

In Fig. 2, we plot \mdos\ against \lum\ for the VC, WP, and SP samples
along with a solid line representing the power law derived from our
isolated galaxies sample (Eq. 1).  We have analyzed the residuals of
each subsample with respect to the isolated galaxies template, and
none of the three subsamples can be distinguished from the isolated
galaxy template. In Table 1, we summarize the median values of the
residuals and the semi-interquartile distances which measure the
dispersion of the distribution.  The deviation of the residuals from
zero value are statistically negligible, and this fact indicates that
neither the optical luminosity {\em nor the molecular gas content} as
measured in CO are significantly affected by tidal interactions. This
contradicts the previous reports of enhanced molecular gas
content among strongly interacting pairs (Braine \& Combes 1993,
Combes et al 1994).  Virgo spirals show a gas content similar to
isolated galaxies as has been previously suggested (see Boselli 1994 
and references therein).

The relation between \fir\ and \mdos\ is plotted in Fig. 3 for the
three subsamples along with the power law relation obtained from the
isolated galaxy sample.  While the Virgo spirals and WP sample
galaxies follow the same \fir-\mdos\ relation as the isolated galaxy
sample, the SP sample galaxies show a clear deviation -- enhanced FIR
emission for given molecular gas content, already known in the
literature as increase ``star formation efficiency''
(L$_{FIR}$/M$_{H_2}$ ratio -- e.g. Sanders et al. 1991).  The FIR
luminosity of the SP sample galaxies lies mostly within the range
covered by the isolated galaxy sample, but it extends to higher FIR
luminosity and has a higher mean FIR luminosity than the isolated
sample, requiring some extrapolation of Eqs. 1 \& 2.  We note however
that the ranges of \mdos\ and \lum\ covered by the SP and isolated
comparison sample are the same, clearly indicating that the FIR
emission is enhanced among the SP sample galaxies.  We have improved
the understanding of L$_{FIR}$/M$_{H_2}$ ratio by proper accounting
the strong non-linear luminosity dependence while previous studies of
L$_{FIR}$/M$_{H_2}$ ratio were performed with a simple assumption of
no luminosity dependence and despite claims of enhanced molecular gas
content in some cases.

Since we have shown that H$_2$ content is independent of tidal
disruption or environment, any deviations from the template will
indicate changes in the FIR emission.  In Table 2 we summarize the
statistics and probability of association in L$_{FIR}$/M$_{H_2}$
distribution for the VC, WP, and SP samples with respect to the
isolated galaxy sample. A zero value for the probability indicates
that the compared distributions are statistically different from the
isolated galaxy sample.  Also included in Table 2 are the median and
the semi--interquartile distance of the log[\fir/\mdos] for all the
samples as obtained from the cumulative distribution functions.  All
performed statistical tests indicate large differences in the
distribution of the SP galaxies with respect to the isolated galaxies,
as it was already clearly shown in Fig. 3.  The VC galaxies show a
$L_{FIR}$--$M_{H_2}$ distribution very similar to the isolated
galaxies as indicated by the similarity of their median values and by
high statistical likelihood.

An evident result is that the star formation efficiency in the SP
sample is enhanced on average by a factor 3.5 with respect to the
isolated galaxy sample, consistent with the results in the literature
(\cite {s86}, Solomon \& Sage 1988,\cite {ys91}), even when various
non-linear dependences are properly accounted (the probability of the
SP sample having the same distribution as the isolated sample is
zero).  Our new analysis also removes any uncertainty regarding
variations in $M_{H_2}$ as function of interaction strength (thus
affecting L$_{FIR}$/M$_{H_2}$ ratio) since no such variation is found.
The weakly interacting systems show some marginal excess in SFE (30\%,
well within one sigma).  This can be understood since observational
separation between weak and strong interactions is not always
straightforward, and some contamination exists between the two
samples.

\subsection{Barred and unbarred galaxies}

Out of 196 galaxies in our sample with morphological classification
listed in NED, 117 are classified as barred (38 isolated, 28 WP, 13
SP, and 38 VC) and 79 as unbarred (30 isolated, 15 WP, 14 SP, and 20
VC). The same analysis is now performed for the barred versus unbarred
galaxies in order to test the previous claims that enhance star
forming activity is associated with barred spirals (e.g., \cite{y92}).

In Fig. 4a, we plot L$_B$ against M$_{H_2}$ for the barred and
unbarred galaxies, and no clear difference is seen between them,
indicating that the total molecular mass is independent of their
barred morphology. In fact L$_B$ distributions of barred and unbarred
galaxies are found indistinguishable by different statistical tests
applied, and the same conclusion is drawn for their \mdos\
distributions also ($<log(L_B)>_{bar}$ = 10.231, $<log(L_B)>_{unbar}$
= 10.110; $<log[M_{H_2}]>_{bar}$ = 9.195, $<log[M_{H_2}]>_{unbar}$ =
9.110).

We have also explored for any differences in L$_{FIR}$--$M_{H_2}$
distribution as shown in Fig. 4b.  Again the data overlap, and the FIR
emission of the barred and unbarred galaxies is found
indistinguishable by the statistical tests performed
($<log(L_{FIR})>_{bar}$ = 9.691, $<log(L_{FIR})>_{unbar}$ = 9.687).
Thus the star formation efficiency seems independent of the barred
morphology, as shown in the histogram of the ratio $L_{FIR}/$$M_{H_2}$
in Fig. 5.  The median values that are respectively
$[L_{FIR}/M_{H_2}]_{unbarred}$ = 0.69 ($\sigma$ = 1.59) and
$[L_{FIR}/M_{H_2}]_{barred}$ = 0.76 ($\sigma$ = 1.58).  Since bar
morphology and starburst activity are sometimes associated with tidal
interactions, this result may indicate that the lifetime of a stellar
bar, if transient, is much longer than the time scale for starburst.

\section{Conclusions}

Through the study of a carefully selected sample of isolated spiral
galaxies, we have clearly established that the two important global
physical quantities for tracing star forming activities, $L_{FIR}$ and
$M_{H_2}$, have non-linear dependence on another commonly cited global
quantity $L_B$.  Furthermore we show that simple power law relations
can effectively describe these non-linear relations for spiral
galaxies spanning four orders of magnitude in FIR and \mdos\ and
nearly three orders of magnitude in $L_B$.  These non-linear relations
are critically important in comparing different samples of galaxies
since a simple normalization and the resulting residual dependence
imply that samples consisting of the more luminous galaxies would show
intrinsically larger ratios.  Using the non-linear template relations
obtained from the isolated galaxy sample, we have re-analyzed previous
claims of enhanced molecular gas content and star formation through
tidal interactions.  We have also examined the effect of cluster
environment and presence of stellar bars on the $H_2$ content and any
enhancement of star formation.

In the frame defined by the optical luminosity, molecular gas content,
and far infrared luminosity, we found that gravitational interaction
affects only the FIR emission.  We found no evidence for enhanced
molecular gas content among the interacting galaxies, and we conclude
that the previous such claims are due to the use of size normalization
using $L_B$ without properly accounting of the non-linear dependence.
We do confirm that enhanced FIR emission (star formation) by tidal
interaction is real.  We found no evidence for enhanced star formation
among barred galaxies in comparison with the unbarred galaxies.

\acknowledgements

The authors thank N. Scoville, P. Solomon and J. Sulentic for useful
discussions.  JP, AO and LV-M are partially supported by DGICYT
(Spain) Grant PB93-0159 and Junta de Andaluc\'{\i}a (Spain).  MSY is
supported by the NRAO Jansky Research Fellowship.

\clearpage

\figcaption[fig1.ps]
{The dependence of \lum\ and \fir\ on \mdos\ for the isolated galaxy
sample. The solid lines correspond to the fitted power law given in
Eq. 1.}

\figcaption[fig2.ps]
{The dependence of \lum\ on \mdos\ for the three samples of
interacting galaxies. The plotted line corresponds to the template
power law derived from the isolated galaxies.}

\figcaption[fig3.ps]
{The dependence of \fir\ on \mdos\ for the interacting galaxy sample.}

\figcaption[fig4.ps]
{The dependence of \lum\ and \fir\ on \mdos\ for all galaxies in our
sample, separated by barred and unbarred morphology.}

\figcaption[fig5.ps]
{Histogram of the star formation efficiency measured by the ratio
$[L_{FIR}/M_{H_2}]$ for our full sample, divided into barred and
unbarred galaxies.}

\clearpage

\begin{table*}
\begin{center}
\begin{tabular}{lll}
\hline
 \multicolumn{1}{c}{Sample} & \multicolumn{1}{c}{Median} &
Q$^{a}$\\
\hline
Isolated& 0.043& 0.17\\
W. Per.& 0.204& 0.21\\
S. Per.& 0.101& 0.27\\
Virgo& 0.197& 0.14\\
\hline
\end{tabular}
\end{center}
\tablenotetext{a}{
Semi-interquartile  distance. For a normal distribution $\sigma$ 
= 3/2 Q. }
\tablenum{1}
\caption{Statistical parameters of the residuals relative to the
\mdos--\lum\ template obtained from isolated galaxies.}

\end{table*}

\clearpage

\begin{table*}
\begin{center}
\begin{tabular}{lllll}
\hline
 & \multicolumn{1}{c}{Isolated} & \multicolumn{1}{c}{Virgo} &
\multicolumn{1}{c}{WP} & \multicolumn{1}{c}{SP} \\
\hline
logrank statistic  & \pepe & 1.103 & 2.193 & 5.769 \\
Probability        & \pepe & 0.270 & 0.028 & 0.000 \\
\hline
Peto--Prentice     & \pepe & 0.191 & 1.889 & 6.457 \\
Probability        & \pepe & 0.848 & 0.059 & 0.000 \\
\hline
Median             & 0.633 & 0.583 & 0.737 & 1.171 \\
Semi-interquartile & 0.287 & 0.404 & 0.279 & 0.184 \\
\hline
\end{tabular}
\end{center}
\tablenum{2}
\caption{Statistical parameters of log[\fir/\mdos] distribution for 
the studied samples as compared with the isolated galaxies 
template (from Astronomy Survival Analysis package).}

\end{table*}
\clearpage
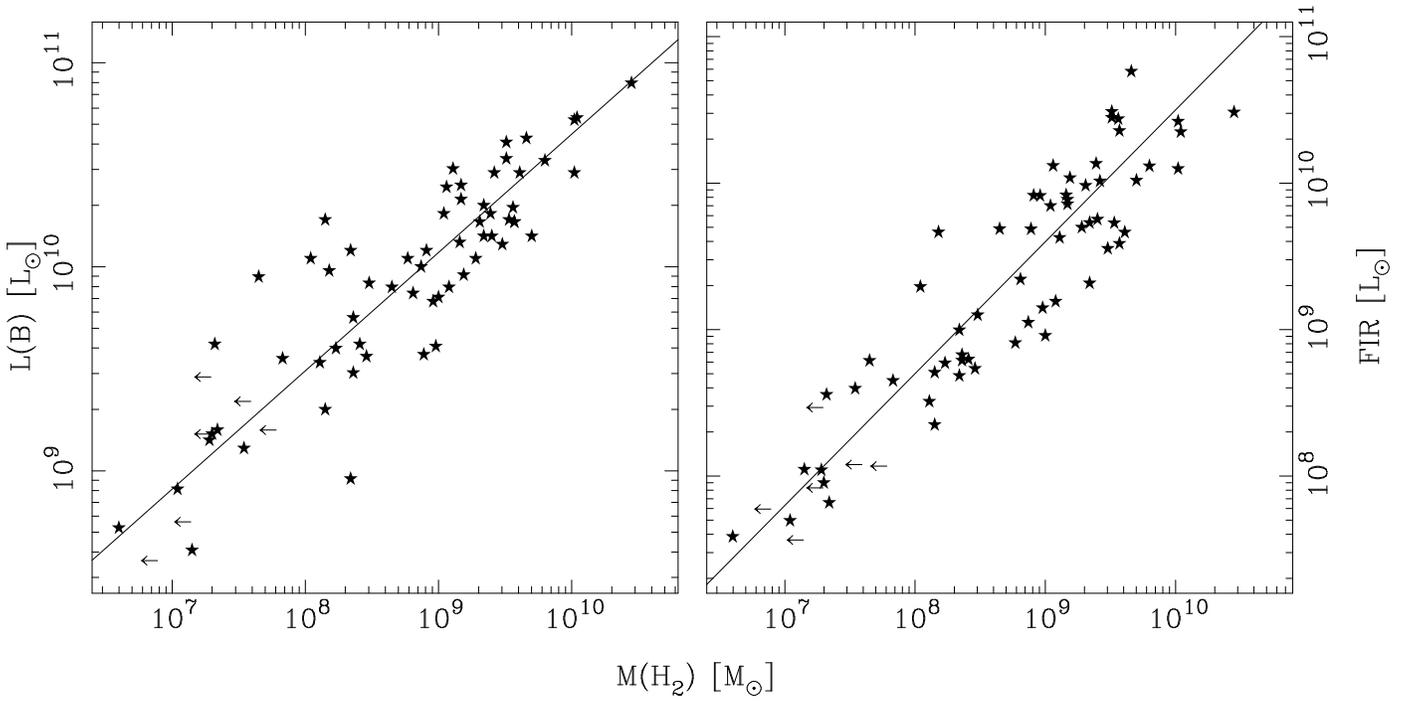
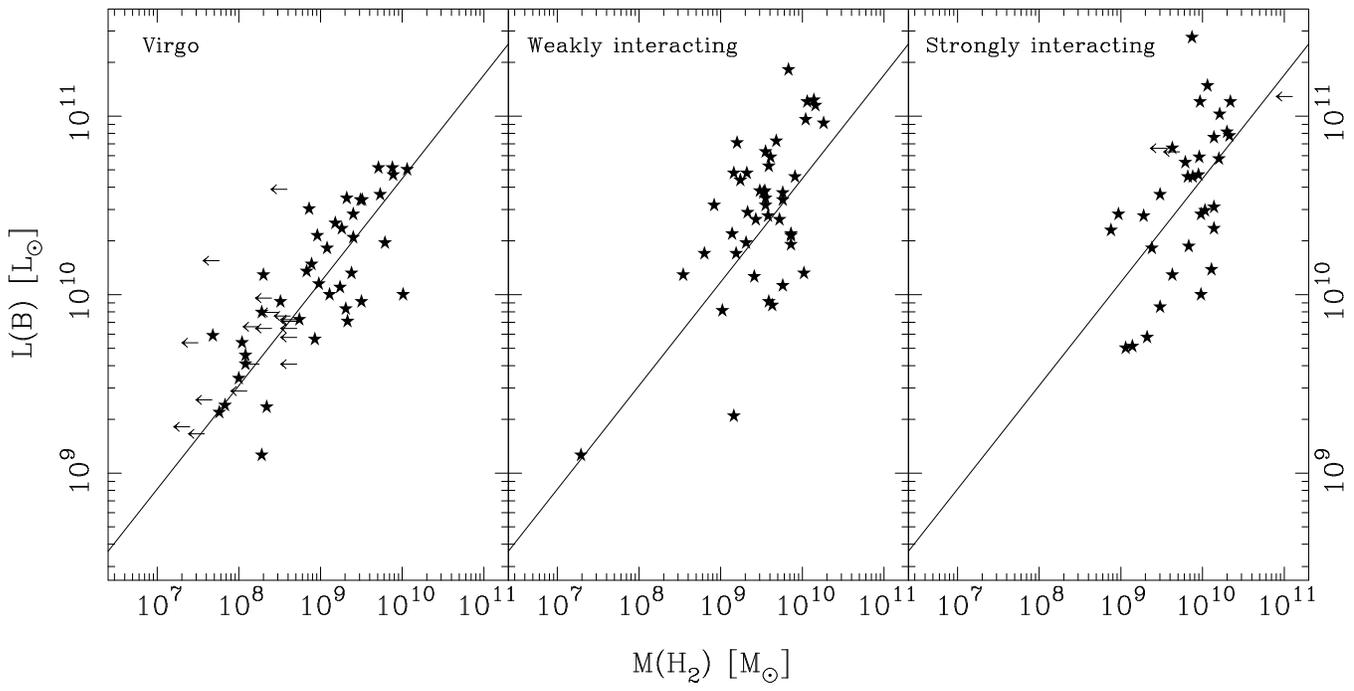
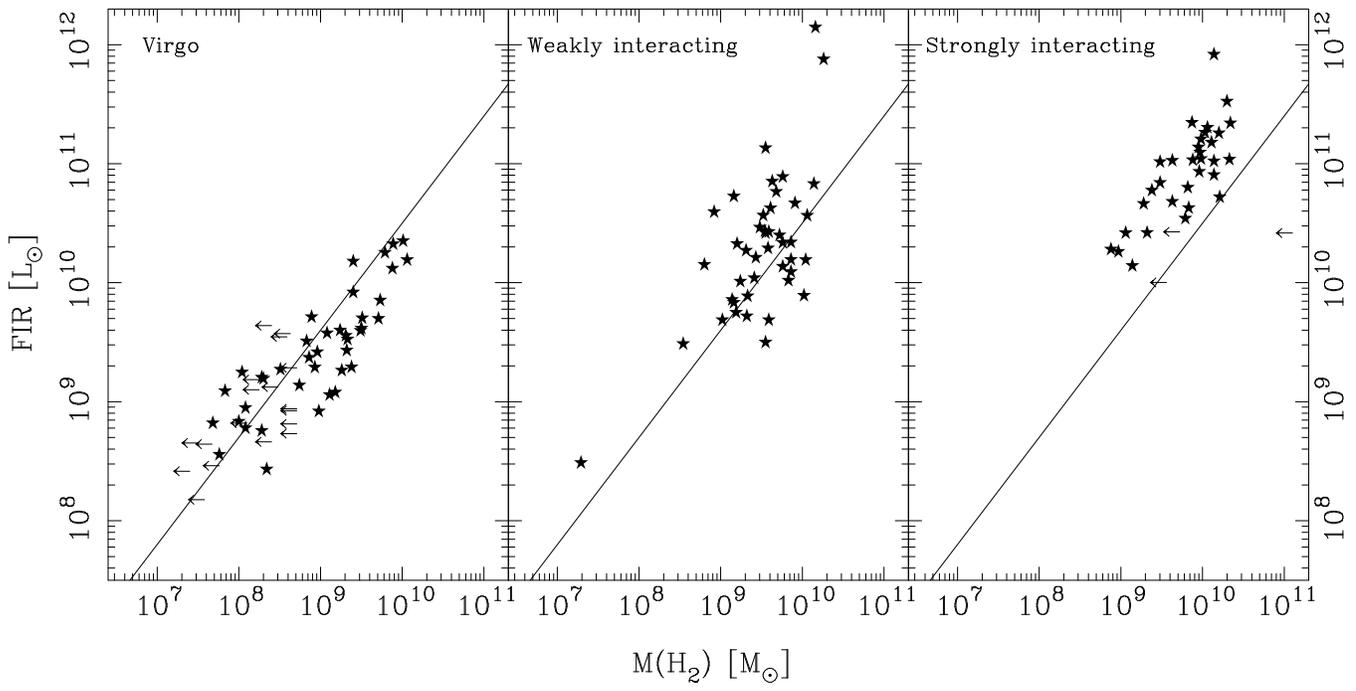
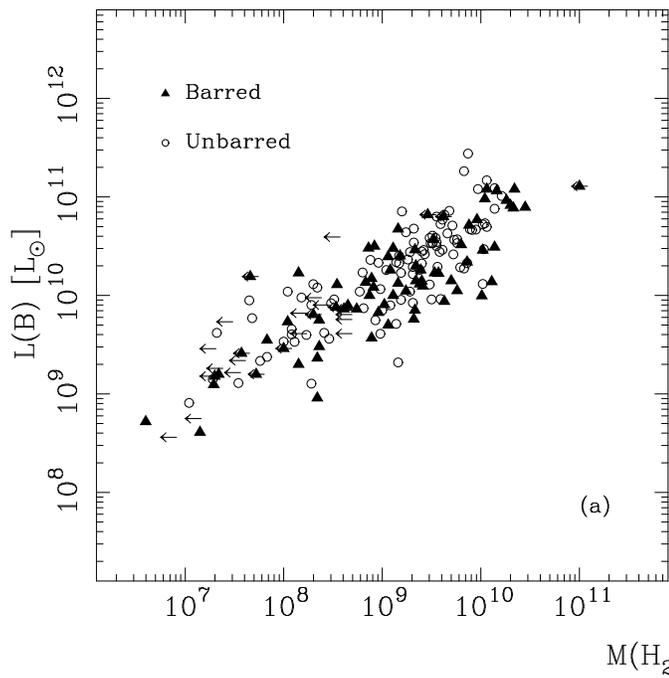
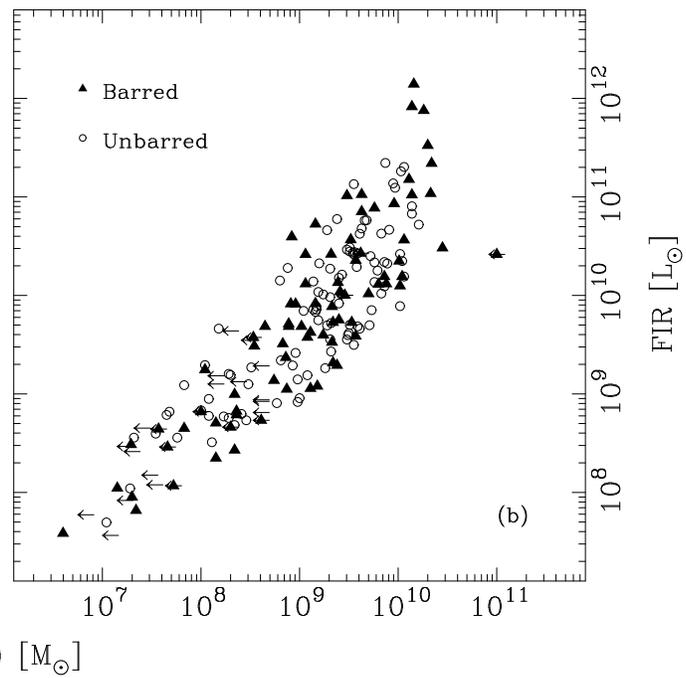
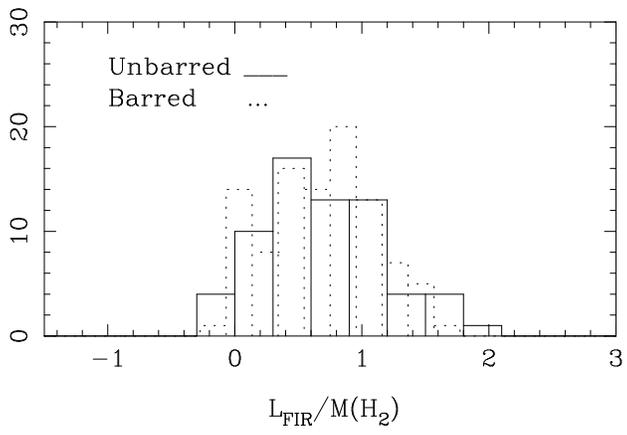
\end{document}